\documentclass[twocolumn,english,pratwocolumn]{revtex4}
\usepackage[latin9]{inputenc}
\usepackage{amsmath}
\usepackage{amssymb}
\usepackage{graphicx}
\usepackage{esint}

\makeatletter

\providecommand{\tabularnewline}{\\}
\newcommand{\lyxdot}{.}

\@ifundefined{textcolor}{}
{%
 \definecolor{BLACK}{gray}{0}
 \definecolor{WHITE}{gray}{1}
 \definecolor{RED}{rgb}{1,0,0}
 \definecolor{GREEN}{rgb}{0,1,0}
 \definecolor{BLUE}{rgb}{0,0,1}
 \definecolor{CYAN}{cmyk}{1,0,0,0}
 \definecolor{MAGENTA}{cmyk}{0,1,0,0}
 \definecolor{YELLOW}{cmyk}{0,0,1,0}
}

\usepackage{babel}

\makeatother

\usepackage{babel}
\begin{document}

\title{Decoherence of Macroscopic Objects from Relativistic Effect}

\author{G. H. Dong$^{1,2}$, Y. H. Ma$^{1,2}$, J. F. Chen$^{1,2}$, Xin
Wang$^{1}$ and C. P. Sun$^{1,2}$}
\email{cpsun@csrc.ac.cn}

\address{$^{1}$Beijing Computational Science Research Center, Beijing 100193,
China \\
 $^{2}$Graduate School of China Academy of Engineering Physics, Beijing
100193, China}
\begin{abstract}
We study how the decoherence of macroscopic objects is induced intrisinically
by relativistic effect. With the degree of freedom of center of mass
(CM) characterizing the collective quantum state of a macroscopic
object (MO), it is found that a MO consisting of $N$ particles can
decohere with time scale no more than $\sqrt{N}^{-1}$. Here, the
special relativity can induce the coupling of the collective motion
mode and the relative motion modes in an order of $1/c^{2}$, which
intrinsically results in the above minimum decoherence.
\end{abstract}

\pacs{03.30.+p, 03.65.Yz, 04.20.Cv}
\maketitle

\section{Introduction}

Quantum superposition lies in the heart of both quantum mechanics
\cite{Griffiths,Sakurai} and the current quantum technologies such
as quantum communication and computation \cite{Nielsen,L-M-Duan}.
Any superposition of different states whose evolution is governed
by Schrödinger's equation \cite{Schr=00003D0000F6dinger equation}
still satisfies the same evolution equation and remains valid in quantum
world. Upon one measurement, Born's rule determines the probability
of one definite outcome \cite{Born rule}. Another feature of interest
in quantum mechanics is the quantum coherence, which is depicted by
superposition. For instance, the fringes in the interference experiment
of electrons show the coherence of electron state in different paths.
However, interaction between a physical system with its environment
may ruin this coherence. This environment-induced process, known as
decoherence or dissipation \cite{open quantum systems}, destroys
the coherence of the system, i.e., suppresses strongly the interference
of states of the system or dissipates its energy, and singles out
a set of states which behave like classical states \cite{Physics-today-Zurek,Zure-Rev-Mod-Phys}.
In the quantum theory of measurement where the system, apparatus and
environment are all treated as quantum objects (governed by Schrödinger's
equation), decoherence is proposed to interpret the outcome of measurements
without the collapse of wave packets \cite{Zure-Rev-Mod-Phys,ZurekMeasurement1981 }.
A number of models of environment which do not dissipate energy of
the system but contribute to its decoherence have been studied in
the past decades. For example, the environment can be chosen as a
ring of spin $1/2$ \cite{HTQuan,spin}, a reservoir of harmonic oscillators
\cite{open quantum systems,Z-Sun-Euro-Phys-J-D-(2008)} and many other
external environments \cite{Joos-and-Zeh}.

It seems that quantum theory, while tested thoroughly at microscopic
level, is somehow counter-intuitive in the macroscopic domain. In
Schrödinger's well-known gedanken experiment \cite{Schr=00003D0000F6dinger's cat},
a cat, which is described as a macroscopic object (MO), is in a superposition
state of alive and dead which has never been observed in the classical
world. In our daily life, the cat is either alive or dead with the
same chance but not in the superposition state. In fact, decoherence
plays an important role in this transition from quantum to classical
world \cite{Physics-today-Zurek,Zure-Rev-Mod-Phys}. In the universe,
isolated systems barely exist, especially the MOs, which must interact
with environments (with a large scale of degrees of freedom). Generally
speaking, the larger the scale of system is, the faster it decoheres
\cite{M. S RMP}.

As we stated above, various interactions will lead to this quantum-classical
transition phenomena. In this paper, we focus on its intrinsic origin
which results in the minimum decoherence even in the absence of any
usual environments. It has been found that the collective mode of
MOs can be coupled with its inner motion modes \cite{C-P-Sun-E-P-J-D,Arxiv-9910140}.
C. Carazza has studied the decoherence effect of the collective variable
for free quasi-relativistic particles \cite{Arxiv-9910140}. Nevertheless,
for macroscopic objects, a more reasonable scenario should take the
interaction between particles into account, since it is the coupling
between particles that bound them into a macroscopic object. Recently,
Igor Pikovski et al. has also studied decoherence due to gravitational
time dilation in 2015 \cite{Igor Nat phy 2015}. They phenomenologically
thought that the internal movement energy of the system can contribute
to its total mass, and the center of mass (CM) was coupled to the
internal movement due to the general relativistic effect.

In this paper, we revisit the effect of the special relativity on
the decoherence of collective mode of macroscopic objects. We study
a ring of relativistic particles with the nearest-neighbouring interaction,
and under some transformation we find there exist interactions between
the CM and the internal degrees of freedom. Then we look into the
decoherence of CM motion after time evolution and obtain the decoherence
time $\tau\sim\left(3\sqrt{N}\left|\triangle E_{1,2}\right|\omega/2Mc^{2}\right)^{-1}$where
$N$ is the particle number, $\triangle E_{1,2}$ is the energy difference
of two initial state, $\omega$ is the coupling strength and $M$
is the total mass of CM.

The remainder of this paper is organized as follows. In Sec. II, we
describe the relativistic macroscopic system consisting of $N$ particles
and obtain its effective Hamiltonian with lowest relativistic effect
to depict the decoherence of CM motion. In Sec. III (IV), we choose
the initial state as a product state of the superposition states of
CM momentum (coherent states) and ground states of simple oscillators,
and figure out the time evolution of this state. Then we obtain the
reduced density matrix and analyze its decoherence. In Sec. V, we
study the decoherence of free particles with the same initial state
and compare it with the outcomes in Sec. III and IV. The conclusions
and discussions are in Sec. VI.

\section{The quasi-relativistic macroscopic object}

In this section, we start from a relativistic MO composed of $N$
particles, each of which obeys the Dirac equation, with the nearest-neighbouring
interaction being considered. It is well-known from the special relativity
theory that the energy modification is introduced in the classical
kinetic part of one particle. In special relativistic quantum theory,
the Dirac Hamiltonian of a free fermion reads \cite{Field theory}
\[
H_{0}=\beta mc^{2}+c\overrightarrow{\alpha}\cdot\overrightarrow{p},
\]
where $p$ and $m$ are the momentum operator and mass of the particle
respectively, and
\[
\beta=\left(\begin{array}{cc}
I & 0\\
0 & -I
\end{array}\right),\alpha_{i}=\left(\begin{array}{cc}
0 & \sigma_{i}\\
\sigma_{i} & 0
\end{array}\right),(i=1,2,3)
\]
are the Dirac matrices and $c$ denotes the light velocity. It is
well-known that there are four eigenvectors where two correspond to
positive energy $\sqrt{p^{2}c^{2}+m^{2}c^{4}}$ and the others correspond
to negative energy $-\sqrt{p^{2}c^{2}+m^{2}c^{4}}$. In other words,
one can diagonalize this Hamiltonian in block with an unitary transformation
$U=\exp\left[\beta\overrightarrow{\alpha}\cdot\overrightarrow{p}\arctan\left(p/mc\right)/2p\right]$
\cite{foldy} as $H_{0}^{'}=UH_{0}U^{\dagger}$, i.e.,
\begin{align*}
H_{0}^{'} & =\beta\sqrt{p^{2}c^{2}+m^{2}c^{4}}.
\end{align*}
The positive and negative energy spaces are separated. In the following
we will focus on the positive energy part and consider the lowest-order
relativistic correction of non-relativistic particles. Actually the
above argument is carried out for Dirac particles. For scalar particles,
one can also obtain the mass-energy relation with Klein-Gordon equation.

Thus the total Hamiltonian of the relativistic MO is
\begin{equation}
H=\sum_{i}\sqrt{p_{i}^{2}c^{2}+m^{2}c^{4}}+\frac{1}{2}m\omega^{2}\left(x_{i}-x_{i+1}\right)^{2},\label{eq:whole Hamiltonian}
\end{equation}
where the $N$ particles are of the same mass $m$, $p_{i}$ and $x_{i}$
are the momentum and position operator at site $i$ respectively as
we set the lattice distance $a=1$. The Hamiltonian to the second-order
approximation which contains the lowest-order term with the relativistic
effect becomes
\begin{equation}
H\simeq Nmc^{2}+\sum_{i}^{N}\left[\frac{p_{i}{}^{2}}{2m}-\frac{p_{i}{}^{4}}{8m^{3}c^{2}}+m\omega^{2}\left(x_{i}^{2}-x_{i}x_{i+1}\right)\right].\label{eq:2-order hamiltonian}
\end{equation}

For simplicity, we choose $N$ as an odd number without loss of generality
and take the Fourier transformation

\begin{align}
p_{k}=\sqrt{\frac{1}{N}}\sum_{j=1}^{N}p_{j}e^{-\frac{i2\pi}{N}kj}, & x_{k}=\sqrt{\frac{1}{N}}\sum_{j=1}^{N}x_{j}e^{\frac{i2\pi}{N}kj},\label{eq:Fourier transformation}
\end{align}
to get the normal modes with $p_{k}$ and $x_{k}$, the momentum and
position operators at momentum space respectively. It is worth mentioning
that $p_{k=N}=\sum_{j=1}^{N}p_{j}\exp\left(-i2\pi j\right)/\sqrt{N}=P/\sqrt{N}$,
and $x_{k=N}=\sum_{j=1}^{N}x_{j}\exp\left(i2\pi j\right)/\sqrt{N}=\sqrt{N}X$
are the momentum and position operators of the CM respectively. That
is to say, the N-th mode describes the motion of CM while the other
N-1 modes describe the internal relative motion. Therefore, we introduce
the CM and relative coordinates in the relativistic Hamiltonian by
this Fourier transformation. In this center of mass reference frame,
the Hamiltonian in low-energy limit becomes,

\begin{align}
H & \simeq\frac{P^{2}}{2M}-\frac{P^{4}}{8M^{3}c^{2}}+\sum_{k=1}^{N-1}\frac{p_{k}p_{-k}}{2m}\nonumber \\
 & +2m\omega^{2}\sum_{k=1}^{N-1}x_{k}x_{-k}\sin^{2}\left(\frac{\pi}{N}k\right)-\frac{3P^{2}}{2M^{2}c^{2}}\sum_{k=1}^{N-1}\frac{p_{k}p_{-k}}{2m}\nonumber \\
 & -\frac{P}{2m^{2}Mc^{2}\sqrt{N}}\sum_{k_{1},k_{2},k_{3}=1}^{N-1}p_{k_{1}}p_{k_{2}}p_{k_{3}}\delta_{k_{1}+k_{2}+k_{3},qN}\nonumber \\
 & -\frac{1}{8m^{2}Mc^{2}}\sum_{k_{1},k_{2},k_{3},k_{4}=1}^{N-1}p_{k_{1}}p_{k_{2}}p_{k_{3}}p_{k_{4}}\delta_{k_{1}+k_{2}+k_{3},qN},\label{eq:3-order couple neglected}
\end{align}
where $\delta_{k_{1}+k_{2}+k_{3},qN}$ for $q$ being one of nonzero
integers, is the Kronecker Delta function. It can be seen in Eq. (\ref{eq:3-order couple neglected})
that the first two terms describe Hamiltonian of CM while the third
and fourth terms the relative motions. Obviously the last three terms
characterize the interaction between CM and relative motion and the
higher order correction.

We then treat the CM system as ``system'' and the relative motion
system as ``internal environment'' and the whole Hilbert space is
the product of two subsystems $\mathcal{H}=\mathcal{H}_{S}\bigotimes\mathcal{H}_{E}$.
Therefore, this division of Hamiltonian in Eq. (\ref{eq:3-order couple neglected})
implies that the motion of the system will be influenced by the environment.
As a consequence
\begin{align}
H & \simeq\frac{P^{2}}{2M}+\sum_{k=1}^{N-1}\left(\frac{p_{k}p_{-k}}{2m}+2m\omega^{2}x_{k}x_{-k}\sin^{2}(\frac{\pi}{N}k)\right)\nonumber \\
 & -\frac{3P^{2}}{2M^{2}c^{2}}\sum_{k=1}^{N-1}\frac{p_{k}p_{-k}}{2m}\nonumber \\
 & =H_{S}+H_{E}+H_{SE},\label{eq:H without diagnal}
\end{align}
where the high order collective terms are neglected. It can be checked
immediately that $[H_{S},H_{SE}]=0$ and $[H_{E},H_{SE}]\neq0$, which
means that evolution governed by Hamiltonian in Eq. (\ref{eq:H without diagnal})
may cause entanglement between the system and the environment. According
to the decoherence theory, this kind of interaction will induce a
transition from the quantum superposition state to the classical statistical
mixture in the system without energy dissipation \cite{open quantum systems,Physics-today-Zurek}.

We have obtained the decoherence model, but there remains another
problem: different modes of relative motions are not independent but
coupled in pairs. Bogoliubov transformation can help us diagonalize
$H_{E}$. This transformation is given by $\left(\begin{array}{cccc}
Q_{k=1} & Q_{k=2} & ... & Q_{k=N-1}\end{array}\right)^{T}=W_{(N-1)\times(N-1)}\left(\begin{array}{cccc}
q_{k=1} & q_{k=2} & ... & q_{k=N-1}\end{array}\right)^{T},$ where $Q$ stands for $P,X$ which are the momentums and the displacements
of the $N-1$ independent relative motions, $q$ for $p,x$ respectively.
The transformation $W$ is
\begin{align}
W & =\sqrt{\frac{1}{2}}\left(\begin{array}{cccccc}
1 & 0 & ... & 0 & 0 & 1\\
0 & 1 & ... & 0 & 1 & 0\\
 &  & ...\\
0 & 0 & 1 & 1 & 0 & 0\\
0 & 0 & -i & i & 0 & 0\\
0 & -i & ... & 0 & i & 0\\
-i & 0 & ... & 0 & 0 & i
\end{array}\right).\label{eq:diagonalize}
\end{align}
For $j\in[1,N-1]$, only two elements are non-vanishing in every row
and column of $W$. Especially, $W_{j,j}=W_{j,N-j}=1/\sqrt{2}$ when
$j\in[1,\left(N-1\right)/2]$ and $W_{j,j}=W_{j,N-j}^{*}=i/\sqrt{2}$
when $j\in[\left(N+1\right)/2,N-1]$. We can check that $W\cdot W^{\dagger}=I_{(N-1)\times(N-1)}$,
and

\begin{align}
\sum_{k=1}^{N-1}q_{k}q_{-k} & =\sum_{k=1}^{N-1}Q_{k}^{2}.
\end{align}
Finally, the diagonalized Hamiltonian becomes $(\omega_{k}=2\omega\sin(\pi k/N))$
\begin{align}
H & =\frac{P^{2}}{2M}+\sum_{k=1}^{N-1}\left(\frac{P_{k}^{2}}{2m}+\frac{m}{2}\omega_{k}^{2}X_{k}^{2}\right)-\frac{3P^{2}}{2M^{2}c^{2}}\sum_{k=1}^{N-1}\frac{P_{k}^{2}}{2m}\nonumber \\
 & =\frac{P^{2}}{2M}+\sum_{k=1}^{N-1}H_{E,k}+H_{I,k}(P).\label{eq:Diaonalized Hamiltonian}
\end{align}
Now we know that there is only kinetic term in system Hamiltonian,
the environment contains $N-1$ modes of simple harmonic oscillator
with $N-1$ eigenfrequencies and the momentum of CM couples with all
the relative motion modes.

\section{\label{sec:Initial-condition,-time} decoherence dynamics}

In the previous section, we obtained the Hamiltonian of the decoherence
shown in Eq. (\ref{eq:Diaonalized Hamiltonian}). While Hamiltonian
in Eq. (\ref{eq:Diaonalized Hamiltonian}) governs the time evolution
of the total system, the evolution equation of the reduced density
matrix of the system is quantum master equation, which is not unitary
due to the interaction with its environment \cite{open quantum systems}.
According to Born's rule, the diagonal terms of the reduced density
matrix describe the probabilities of getting some outcome in one measurement,
while the off-diagonal ones characterize the interference of different
quantum states and show the coherence properties of this system. When
the coherence of the system decays with time while the probability
terms remain stable, the system undergoes an transition from quantum
to classical, i.e., decoherence \cite{Zure-Rev-Mod-Phys,M. S RMP}.

We can see that, in the Hamiltonian (\ref{eq:Diaonalized Hamiltonian}),
the relative motion are N-1 simple harmonic oscillations and the minimal
energy difference between two neighbouring levels in large N limit
is $2\hbar\omega\pi/N$. Assuming the temperature of environment is
too low to excite the relative mode, i.e., $2\hbar\omega\pi/N\gg k_{B}T$,
all the relative motion stay in the ground state. Then we choose the
initial state as $\left|\varphi(0)\right\rangle =(\left|P_{1}\right\rangle +\left|P_{2}\right\rangle )/\sqrt{2}\bigotimes\prod_{k=1}^{N-1}\left|0\right\rangle _{k}$
where $\left|P_{i}\right\rangle (i=1,2)$ is the eigenstate of $P$
with eigenvalue $P_{i}$ and $\left|0\right\rangle _{k}$ is the ground
state of $k$-th mode of relative motion Hamiltonian. The total density
matrix at time t evolves as
\begin{align}
\rho(t) & =e^{-iHt/\hbar}\left|\varphi(0)\right\rangle \left\langle \varphi(0)\right|e^{iHt/\hbar}.
\end{align}
The reduced density matrix of the motion of CM is
\begin{align*}
\rho_{c.m.}(t) & =Tr_{E}\rho(t).
\end{align*}
As $\left|P_{1}\right\rangle $ and $\left|P_{2}\right\rangle $ are
the eigenstates of $H_{S}$, the diagonal terms of the local (reduced)
density matrix in basis $\left|P_{i}\right\rangle $ are independent
of time, i.e., $\rho_{S}^{1,1}(t)=\rho_{S}^{1,1}(0)=\rho_{S}^{2,2}(t)=\rho_{S}^{2,2}(0)=1/2$.
This feature indicates that we are dealing with a pure decoherence
process without dissipation. The off-diagonal term reads

\begin{align}
\left|\rho_{c.m.}^{12}(t)\right| & =\frac{1}{2}\prod_{k=1}^{N-1}\left|\left\langle 0\right|e^{iH_{k}(P_{1})t/\hbar}e^{-iH_{k}(P_{2})t/\hbar}\left|0\right\rangle _{k}\right|\nonumber \\
 & =\frac{1}{2}\prod_{k=1}^{N-1}\left|f_{k}(P_{1},P_{2},t)\right|,
\end{align}
where
\begin{align*}
H_{k}(P_{i}) & =H_{E,k}+H_{I,k}(P_{i}),\\
f_{k}(P_{1},P_{2},t) & =\left\langle 0\right|S_{k}(\xi_{1}e^{i\pi})S_{k}(r_{1})S_{k}(r_{2}e^{i\pi})S_{k}(\xi_{2})\left|0\right\rangle _{k},
\end{align*}
where $S\left(r\right)$ is a squeeze operator. Thus the coherence
property of this system is related with the expectation value of four
squeeze operators over the vacuum state (for more details see Appendix
A and B). After some calculation, we obtain
\begin{align}
\lim_{N\gg1}\left|\rho_{c.m.}^{12}(t)\right| & =\frac{1}{2}\exp\left[-\frac{N}{N_{0}}\left(1-J_{0}(4\omega t)\right)\right],\label{eq:N>>1}
\end{align}
where $N_{0}=\left(32M^{2}c^{4}/9\left(\triangle E_{1,2}\right)^{2}\right)$,
$J_{n}(x)$ is the first kind Bessel function and $\triangle E_{1,2}=\left(P_{1}^{2}-P_{2}^{2}\right)/2M$.
One can conclude from Eq. (\ref{eq:N>>1}) that in large $N$ limit
the decoherence function depends on the scale of the system, the energy
difference of initial states, the coupling strength of real particles
and time $t$. One may also find that as $\omega t\ll1$, $J_{0}(4\omega t)\simeq1-4\omega^{2}t^{2}$,
thus the decoherence function becomes

\begin{equation}
\left|\rho_{c.m.}^{12}(t)\right|\simeq\frac{1}{2}\exp\left[-4N\omega^{2}t^{2}/N_{0}\right].\label{small t}
\end{equation}
Then the decoherence time (assuming $N\gg N_{0}$) is
\begin{equation}
\tau\sim\frac{2\sqrt{2}Mc^{2}}{3\sqrt{N}\left|\triangle E_{1,2}\right|\omega}.\label{eq:decoherence time}
\end{equation}
In the long time limit, $\omega t\gg1$, $J_{0}(4\omega t)\simeq0$
and $\left|\rho_{c.m.}^{12}(t)\right|\simeq\exp\left[-N/N_{0}\right]/2$
which indicates that only when $N\gg N_{0}$, $\left|\rho_{S}^{12}(t)\right|\stackrel{\omega t\gg1}{\longrightarrow}0$,
i.e., there is a restriction on the scale of the whole system.

\begin{figure}
\begin{centering}
\includegraphics[scale=0.45]{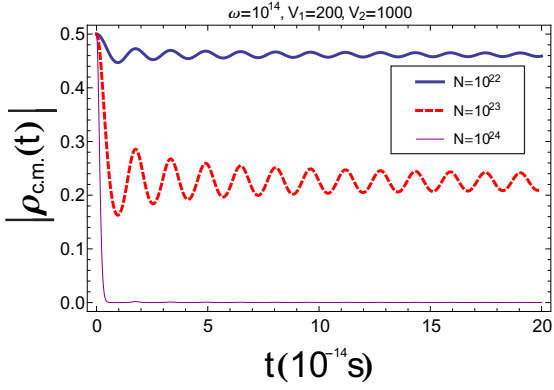}
\par\end{centering}
\caption{\label{fig:The-off-diagonal-matrix}The off-diagonal matrix element
of $\left|\rho_{c.m.}^{12}(t)\right|$ as a function of $t$ with
different N: $N=10^{22}<N_{0}$ (blue,solid),$N=10^{23}\simeq N_{0}$
(red, dashed),$N=10^{24}\gg N_{0}$ (purple,thin). }
\end{figure}

Eqs. (\ref{small t},\ref{eq:decoherence time}) are the main results
in our paper. The decoherence process of CM depends on the scale of
the system, the interaction strength of real particles and the difference
of the initial kinetic energy. The larger the system is, the faster
it decoheres. Tab. (\ref{tab:Decoherence-time-and}) shows a list
of decoherence results with respect to systems of different magnitudes
(e.g., the universe, the earth, person and $C_{60}$). For simplicity,
we have assumed that all these systems are composed of only carbon
atoms. As an example, we consider $C_{60}$ molecules with coupling
strength $\omega\simeq10^{14}$Hz \cite{c=00003D00003Dcstrength}.
With the two superposed initial velocities of this carbon ring as
$200m/s$ and $1000m/s$ \cite{C60-nature}, the lower bound of the
particle number is $N_{0}=1.3\times10^{23}$ which has the magnitude
of Avogadro constant $N_{A}=6.02\times10^{23}$. Thus no decoherence
occurs in a single $C_{60}$ molecure. Fig. (\ref{fig:The-off-diagonal-matrix})
shows $\left|\rho_{c.m.}^{12}(t)\right|$ as a function of $t$ with
different $N$ (assuming that particle number $N$ is changeable in
$C_{60}$). First, when $N\ll N_{0}$, the off-diagonal element $\left|\rho_{c.m.}^{12}(t)\right|$
oscillates around $0.48$ which means the system barely decoheres;
then, as $N\simeq N_{0}$, $\left|\rho_{c.m.}^{12}(t)\right|$ decreases
in the beginning and oscillates around $0.25$ later indicating that
a portion of coherence is retained in the system; at last, for $N\gg N_{0}$,
compared with the above two cases, $\left|\rho_{c.m.}^{12}(t)\right|$
decreases to zero immediately thus the states of the CM in this system
decoheres. And the decoherence time of this MO is about $10^{-14}$s.

It is shown in Eq. (\ref{eq:N>>1}) that the decoherence process depends
on $P_{1}^{2}-P_{2}^{2}$. It seems obscure that when $P_{1}=-P_{2}$
the coherence of CM remains unchanged. In fact, this can be seen in
Eqs. (\ref{eq:3-order couple neglected}, \ref{eq:H without diagnal})
that we drop the term $\sim P\sum_{k_{1},k_{2},k_{3}=1}^{N-1}p_{k_{1}}p_{k_{2}}p_{k_{3}}\delta_{k_{1}+k_{2}+k_{3},N}$
as a higher order term and only keep the one $\sim P^{2}\sum_{k=1}^{N-1}p_{k}p_{-k}$.
This reduction may be related with our results $\propto\left(P_{1}^{2}-P_{2}^{2}\right)^{2}$.
Now it is time to find out the decoherence time as $P_{1}=-P_{2}$.
In this case, the off-diagonal term of the reduced density matrix
is calculated as
\[
\left|\rho_{c.m.}^{12}(t)\right|=\frac{1}{2}\left|\left\langle 0\right|e^{iH(P_{1})t/\hbar}e^{-iH(P_{2})t/\hbar}\left|0\right\rangle \right|,
\]
where
\begin{align*}
H(P_{1}) & =\frac{P_{1}^{2}}{2M}+\sum_{k=1}^{N-1}\frac{P_{k}^{2}}{2m}+\frac{m}{2}\sum_{k=1}^{N-1}X_{k}^{2}\omega_{k}^{2}-\frac{3P_{1}^{2}}{2M^{2}c^{2}}\sum_{k=1}^{N-1}\frac{P_{k}^{2}}{2m}\\
 & -\frac{P_{1}}{2m^{2}Mc^{2}\sqrt{N}}\sum_{k_{1},k_{2},k_{3}=1}^{N-1}W^{\dagger}\left(P_{k_{1}}P_{k_{2}}P_{k_{3}}\right).
\end{align*}
Here $\sum_{k_{1},k_{2},k_{3}=1}^{N-1}W^{\dagger}\left(P_{k_{1}}P_{k_{2}}P_{k_{3}}\right)$
denotes the term $\sum_{k_{1},k_{2},k_{3}=1}^{N-1}p_{k_{1}}p_{k_{2}}p_{k_{3}}\delta_{k_{1}+k_{2}+k_{3},N}$
transformed by $W$ given in Eq. (\ref{eq:diagonalize}). Keeping
terms up to $t^{2}$, one find

\begin{widetext}
\begin{align}
\left|\rho_{c.m.}^{12}(t)\right| & \simeq\frac{1}{2}\left|1+i\frac{t}{\hbar}\left\langle 0\right|H(P_{1})-H(P_{2})\left|0\right\rangle -\frac{t^{2}}{2\hbar^{2}}\left\langle 0\right|\left(H(P_{1})-H(P_{2})\right)^{2}\left|0\right\rangle \right|\nonumber \\
 & =\frac{1}{2}\exp\left(-\frac{t^{2}}{\tau^{2}}-\frac{t^{2}}{\tau^{'2}}\right).
\end{align}

\end{widetext} The decoherence times $\tau$ and $\tau^{'}$ are
obtained which are caused by the $P^{2}$ and $P$ terms of interaction
respectively. What is more important,
\begin{align}
\tau^{-1} & =\frac{9}{32}\frac{\left(V_{1}^{2}-V_{2}^{2}\right)^{2}}{c^{4}}\omega^{2},\nonumber \\
\tau^{'-1} & \simeq\frac{\left(V_{1}-V_{2}\right)^{2}\omega^{2}}{32Mc^{4}}\hbar\omega,\\
\frac{\tau^{'}}{\tau} & \simeq Nm\frac{\left(V_{1}+V_{2}\right)^{2}}{\hbar\omega},\nonumber
\end{align}
where $V_{1}(V_{2})=P_{1}(P_{2})/M$. Here we know that when $P_{1}=-P_{2}$,
the decoherence effect originates from the interaction term $P$ with
time scale $\sim\left(V_{1}-V_{2}\right)^{2}\hbar\omega^{3}/32Mc^{4}$.
Actually, when $P_{1}^{2}\neq P_{2}^{2}$ and in the large system
limit, we also find that the influence of the $P^{2}$ term dominates,
i.e., $\tau\ll\tau^{'}$. In other words, the decoherence process
caused by the interaction term $P^{2}$ is much faster than process
caused by term $P$. And this is why we only keep the interaction
term $P^{2}$ in the very beginning of our paper.

\begin{widetext}

\begin{table}
\begin{centering}
\begin{tabular}{|c|c|c|c|c|}
\hline
 & Universe  & Earth  & Person  & $C_{60}$\tabularnewline
\hline
\hline
initial velocity of CM/$m/s$  & $2.0\times10^{6}$, $1.0\times10^{6}$  & $3.0\times10^{4}$, $2.0\times10^{4}$  & $10$, $5.0$  & $1.0\times10^{3}$, $2.0\times10^{2}$\tabularnewline
\hline
particle number  & $5.0\times10^{78}$  & $2.5\times10^{50}$  & $5.0\times10^{27}$  & $60$\tabularnewline
\hline
particle number bound  & $1.3\times10^{10}$  & $4.0\times10^{17}$  & $2.0\times10^{31}$  & $1.3\times10^{23}$\tabularnewline
\hline
decoherence time/$s$  & $2.0\times10^{-49}$  & $2.0\times10^{-31}$  & $3.0\times10^{-13}$  & $2.4\times10^{-5}$\tabularnewline
\hline
\end{tabular}
\par\end{centering}
\caption{\label{tab:Decoherence-time-and}Decoherence time and particle number
bound of systems of different scale with coupling strength $\omega=10^{14}Hz$. }
\end{table}

\end{widetext}

\section{Decoherence of Cat : superposition of coherent states}

It is well-known that the eigenstates of momentum operators are ideal
quantum states and difficult to prepare in experiments. And more ``classical''
states in quantum mechanics are coherent states, like $\left|\alpha\right\rangle $,
which are generated by applying the displacement operators on the
vacuum states. One important property of the coherent state is that
it satisfies the minimum uncertainty relation. Moreover, coherent
states behave as Gaussian wave packets in both momentum and position
space. In this section, we study the decoherence of a more ``classical''
quantum state, cat state, i.e., the superposition of coherent states.
Here Wigner function in phase space is an useful tool in exploring
the non-classicality of quantum states \cite{ZurekPhysRevD47488(1993)}.

Here we consider that the CM is prepared in a macroscopic superposition
state $\sim\left|\alpha\right\rangle +\left|\beta\right\rangle $
initially, and the relative modes are still in ground states, describing
a non-excited internal enviroment,
\begin{align}
\left|\varPsi(0)\right\rangle  & =\frac{1}{\varXi}\left(\left|\alpha\right\rangle +\left|\beta\right\rangle \right)\bigotimes\prod_{k=1}^{N-1}\left|0\right\rangle _{k},
\end{align}
where $\varXi$ is the normalization factor and $\alpha\left(\beta\right)$
is a complex number. The reduced density matrix of the motion of CM
is

\begin{align}
\rho_{c.m.}(t) & =\frac{1}{\left|\varXi\right|^{2}}\intop\intop dP_{1}dP_{2}\Pi\left(\alpha,\beta,P_{1},P_{2},t\right)\left|P_{1}\right\rangle \left\langle P_{2}\right|,\label{eq:rho-c.m.}
\end{align}
where
\begin{align*}
\Pi\left(\alpha,\beta,P_{1},P_{2},t\right) & =\prod_{k=1}^{N-1}f_{k}(P_{1},P_{2},t)\\
 & *\left(\left\langle P_{1}\mid\alpha\right\rangle +\left\langle P_{1}\mid\beta\right\rangle \right)\left(\left\langle \alpha\mid P_{2}\right\rangle +\left\langle \beta\mid P_{2}\right\rangle \right).
\end{align*}
As illustrated in Sec. \ref{sec:Initial-condition,-time}, the decoherence
function relies on the overlap of two quantum states, $\prod_{k=1}^{N-1}f_{k}(P_{1},P_{2},t)$.
For simplicity, we replace this overlap function with its modulus,
\begin{align*}
\prod_{k=1}^{N-1}f_{k}(P_{1},P_{2},t) & \simeq\prod_{k=1}^{N-1}\left|f_{k}(P_{1},P_{2},t)\right|.
\end{align*}
In momentum representation, the coherent state behaves as a Gaussian
wave packet with packet width $\hbar/2\sigma$ and the mean momentum
$\hbar\Im(\alpha)/\sigma$,
\begin{equation}
\left\langle P\mid\alpha\right\rangle =\left(\frac{2\sigma^{2}}{\pi\hbar^{2}}\right)^{1/4}e^{-\frac{\sigma^{2}}{\hbar^{2}}\left(P-\frac{\hbar}{\sigma}\Im(\alpha)\right)^{2}}e^{-2i\frac{\sigma}{\hbar}\Re(\alpha)P}.\label{eq:coherent state in momentum representation}
\end{equation}
At first glance, it seems difficult to deal with the reduced density
matrix given in Eq. (\ref{eq:rho-c.m.}) as its dimension is infinite.
In order to quantify this decoherence process of CM, we introduce
the quasi-probability distribution, Wigner function, which is defined
as
\begin{equation}
W(p,q)=\frac{1}{\pi\hbar}\int_{-\infty}^{\infty}dye^{i2yp/\hbar}\left\langle q-y\right|\rho\left|q+y\right\rangle .\label{eq:Wigner function}
\end{equation}
Although the Wigner function is a real function, it can not be interpreted
as a probability distribution function since it can be negative. Nevertheless,
if integrating it over $p$ ($q$), one will get the probability distribution
function of $q$ ($p$).

\begin{figure}
\begin{centering}
\includegraphics[scale=0.32]{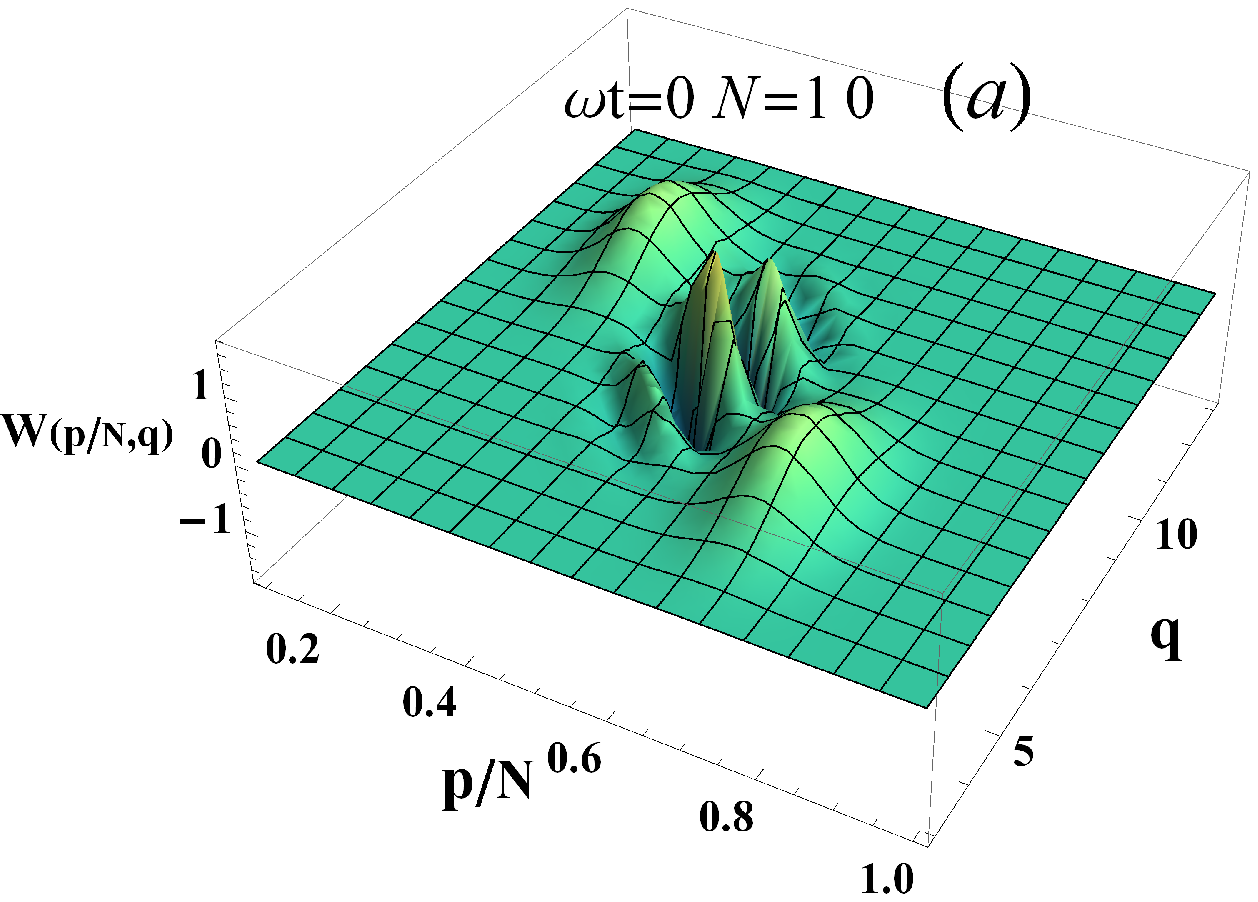}\includegraphics[scale=0.32]{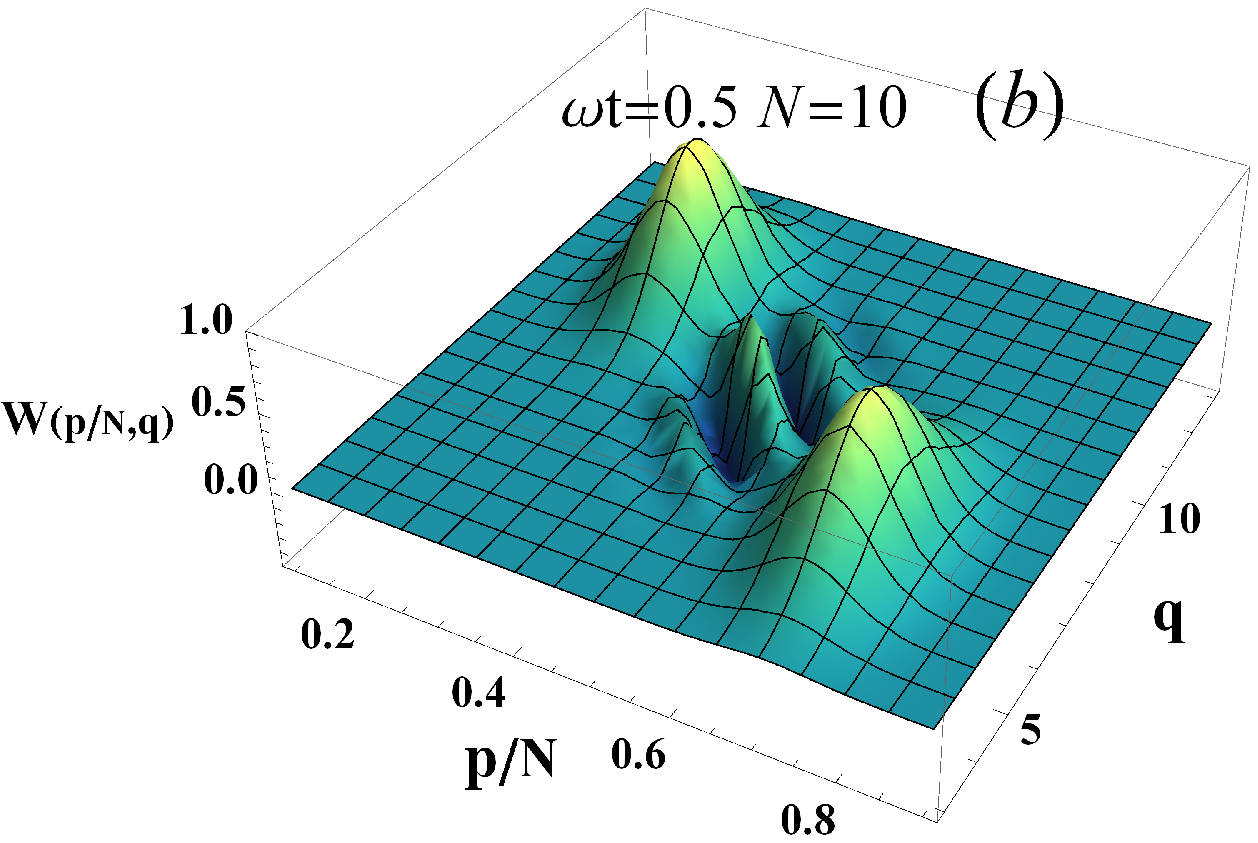}
\par\end{centering}
\begin{centering}
\includegraphics[scale=0.32]{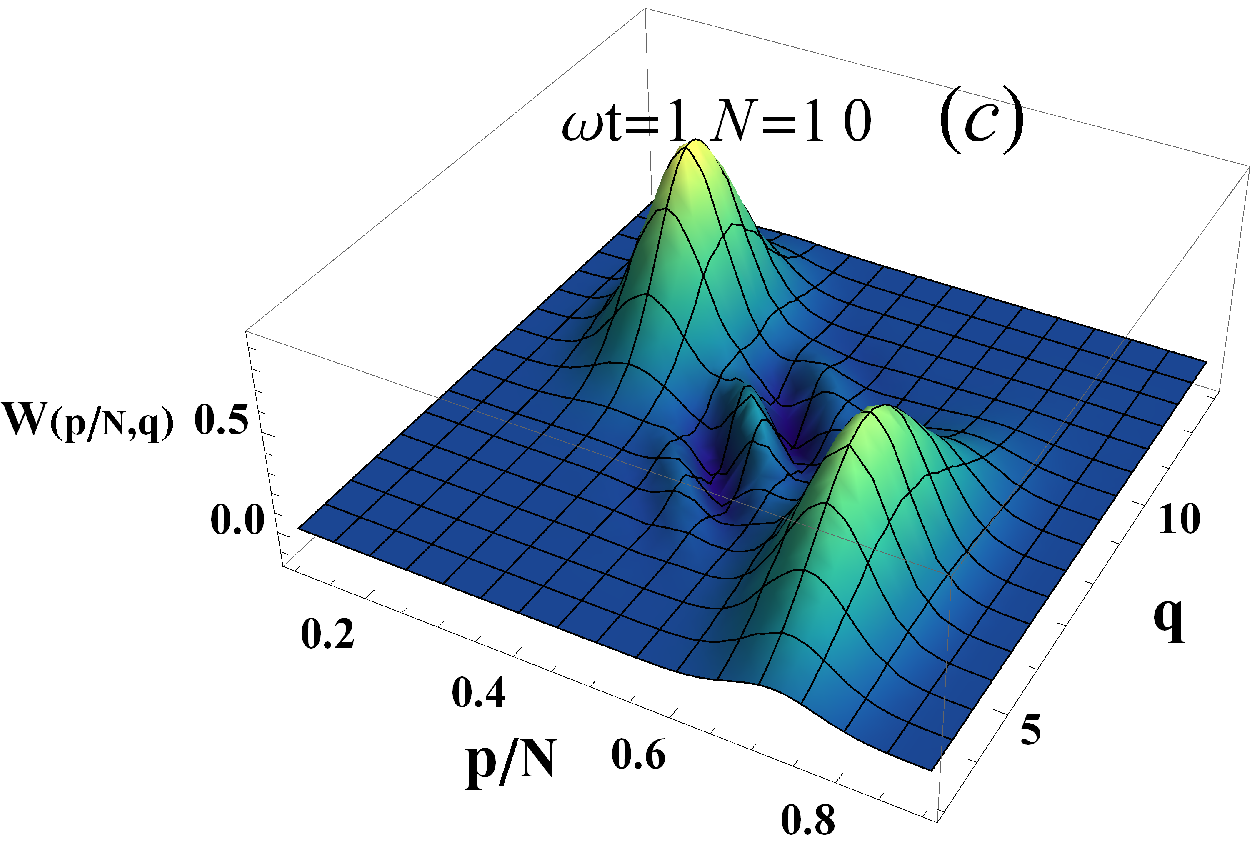}\includegraphics[scale=0.32]{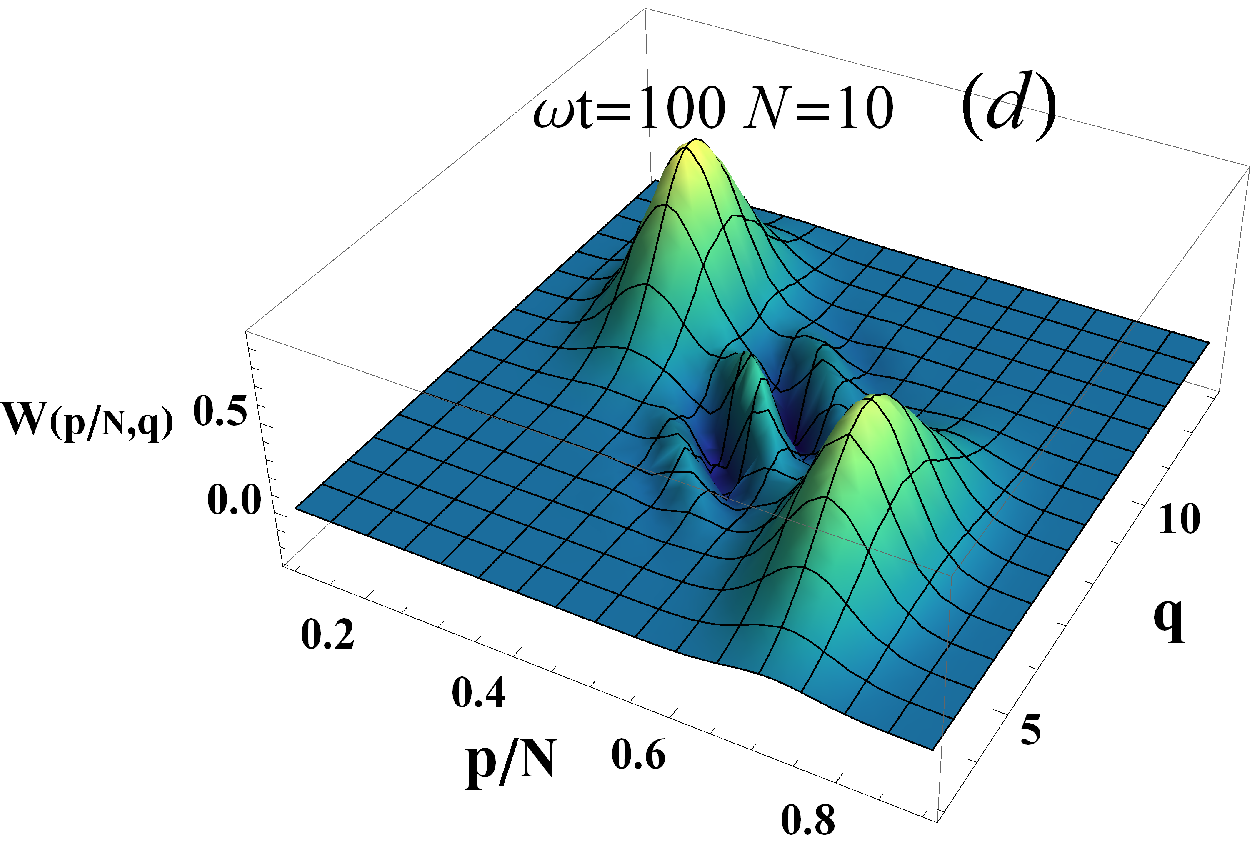}
\par\end{centering}
\centering{}\caption{\label{fig:The-Wigner-function}The Wigner function in phase space
$(p/N,q)$ at different time where we have chosed the natural units
$\hbar=c=1$ and assumed $\sigma=1$, $N=10$, $\gamma=10$, $\alpha=5+3i$
and $\beta=3+7i$: (a) t=0; (b) $\omega t=0.5$; (c) $\omega t=1$;
(d) $\omega t=1$. }
\end{figure}

\begin{widetext}

Inserting Eqs. (\ref{eq:rho-c.m.}, \ref{eq:coherent state in momentum representation})
into Eq. (\ref{eq:Wigner function}), we caculate the Wigner function
of center of mass
\begin{align}
W_{c.m.}(p,q,t) & =W^{\alpha}(p,q,t)+W^{\beta}(p,q,t)+W^{I}(p,q,t),
\end{align}
where

\begin{align}
W^{\alpha}(p,q,t) & =\frac{1}{\left|\varXi\right|^{2}\pi\hbar}e^{-G_{q}^{2}\left(-\Re(\alpha)\right)/2\sigma^{2}}e^{-2\left(\sigma p-\hbar\Im(\alpha)\right)^{2}/\hbar^{2}}e^{-16\gamma\left(1-J_{0}(4\omega t)\right)p^{2}\frac{\sigma^{2}}{4N^{4}\hbar^{2}}\left(1-\frac{G_{q}^{2}\left(-\Re(\alpha)\right)}{\sigma^{2}}\right)},\nonumber \\
W^{\beta}(p,q,t) & =\frac{1}{\left|\varXi\right|^{2}\pi\hbar}e^{-G_{q}^{2}\left(-\Re(\beta)\right)/2\sigma^{2}}e^{-2\left(\sigma p-\hbar\Im(\beta)\right)^{2}/\hbar^{2}}e^{-16\gamma\left(1-J_{0}(4\omega t)\right)p^{2}\frac{\sigma^{2}}{4N^{4}\hbar^{2}}\left(1-\frac{G_{q}^{2}\left(-\Re(\beta)\right)}{\sigma^{2}}\right)},
\end{align}
are the direct Wigner functions contributed by quantum state $\left|\alpha\right\rangle $
and $\left|\beta\right\rangle $ respectively, and

\begin{align*}
W^{I}(p,q,t) & =\frac{1}{\left|\varXi\right|^{2}\pi\hbar}e^{-2\frac{\sigma^{2}}{\hbar^{2}}\left(p+\frac{i\hbar}{2\sigma}\left(\alpha-\beta^{*}\right)\right)^{2}}e^{-\left(\alpha-\beta^{*}\right)^{2}/2-\Im(\alpha)^{2}-\Im(\beta)^{2}}e^{-G_{q}^{2}\left(-\frac{\alpha+\beta^{*}}{2}\right)/2\sigma^{2}}\\
 & *\exp\left[-16\gamma\left(1-J_{0}(4\omega t)\right)p^{2}\frac{\sigma^{2}}{4N^{4}\hbar^{2}}\left(1-\frac{G_{q}^{2}\left(-\frac{\alpha+\beta^{*}}{2}\right)}{\sigma^{2}}\right)\right]+h.c.
\end{align*}

\end{widetext}is the one caused by the interference term $\left|\alpha\right\rangle \left\langle \beta\right|+\left|\beta\right\rangle \left\langle \alpha\right|$,
and we have set $G_{q}(x)\equiv q+2\sigma x$ and $\gamma=9N\hbar^{4}/128\sigma^{4}m^{4}c^{4}$.
In the beginning, the two direct Wigner functions in phase space are
centered at $\left(q=2\sigma\Re\left(\alpha\right),p=\hbar\Im(\alpha)/\sigma\right)$
and $\left(q=2\sigma\Re\left(\beta\right),p=\hbar\Im(\beta)/\sigma\right)$
respectively, which are exactly the mean position and momentum of
two coherent states, while the interference one is approximately centered
at $\left(q=\sigma\left(\Re\left(\alpha\right)+\Re\left(\beta\right)\right),p=\hbar\left(\Im(\alpha)+\Im(\beta)\right)/2\sigma\right)$
(the midpoint of the centers of the two direct terms). Fig. (\ref{fig:The-Wigner-function}a)
shows the Wigner function in phase space at $t=0$, where two packets
correspond to the two coherent states and the oscillations correspond
to the interference term. Since $p$ stands for the momentum of the
whole system (center of mass), $p/N$ indicates the mean momentum
of a single particle. In this case, the two packets are centered at
$(p/N=0.3,q=10)$ and $(p/N=0.7,q=6)$ respectively. Then here comes
the question: How does the total Wigner function evolves with time?

First, let us turn to a mathematical function $\Omega(x,y)=\exp\left[-x^{2}-2(ay-b)^{2}-dy^{2}(1-2x^{2})\right]$
where $a>0,d\geq0$. One can prove that if $2a^{2}\gg d$ and $a^{2}>2b^{2}d$,
the position of its peak is $x=0,y=b/a$, $\Omega(x,y)\mid_{peak}=\Omega(x=0,y=b/a)=\exp\left[-db^{2}/a^{2}\right]$.
Then in our occasion, if $2\gamma\left(1-J_{0}(4\omega t)\right)\ll N^{4}$,
$8\gamma\left(1-J_{0}(4\omega t)\right)\Im(\alpha)^{2}<N^{4}$ and
$8\gamma\left(1-J_{0}(4\omega t)\right)\Im(\beta)^{2}<N^{4}$ (our
choice above meets all these conditions), the positions of the peaks
of three Wigner functions terms remain unchanged with time (see Fig.
(\ref{fig:The-Wigner-function}))
\begin{align}
W^{\alpha}(p,q,t)\mid_{peak} & =\frac{1}{\left|\varXi\right|^{2}\pi\hbar}e^{-4\gamma\left(1-J_{0}(4\omega t)\right)\Im(\alpha)^{2}/N^{4}},\nonumber \\
W^{\beta}(p,q,t)\mid_{peak} & =\frac{1}{\left|\varXi\right|^{2}\pi\hbar}e^{-4\gamma\left(1-J_{0}(4\omega t)\right)\Im(\beta)^{2}/N^{4}},\\
W^{I}(p,q,t)\mid_{peak} & \simeq W^{I}\left(\frac{\hbar}{2\sigma}\Im(\alpha+\beta),\sigma\Re(\alpha+\beta),t\right).\nonumber
\end{align}
It is depicted by the above equation that the peak value of the three
Wigner function terms at time $t$. Since the two direct terms describe
the probability distributions of two coherent states respectively
and the oscillation term describes the interference effect, it sounds
reasonable to quantify the decoherence of the superposition of coherent
states with the peak values of the Wigner function. An useful quantity
introduced by Zurek \cite{ZurekPhysRevD47488(1993)} is the fringe
visibility function
\begin{equation}
F(\alpha,\beta,t)\simeq\frac{1}{2}\frac{W^{I}(p,q,t)\mid_{peak}}{\left(W^{\alpha}(p,q,t)\mid_{peak}W^{\beta}(p,q,t)\mid_{peak}\right)^{1/2}}.
\end{equation}
Obviously, the fringe visibility function describes the decay of the
peak value of the interference term. In other words, it shows the
decoherence of the system which originates from its interaction with
environment. Here, the fringe visibility function gives
\begin{align*}
F(\alpha,\beta,t) & \propto\exp\left[-\gamma\left(1-J_{0}(4\omega t)\right)\left(\Im(\alpha)^{2}-\Im(\beta)^{2}\right)^{2}\right].
\end{align*}

\begin{figure}
\begin{centering}
\includegraphics[scale=0.5]{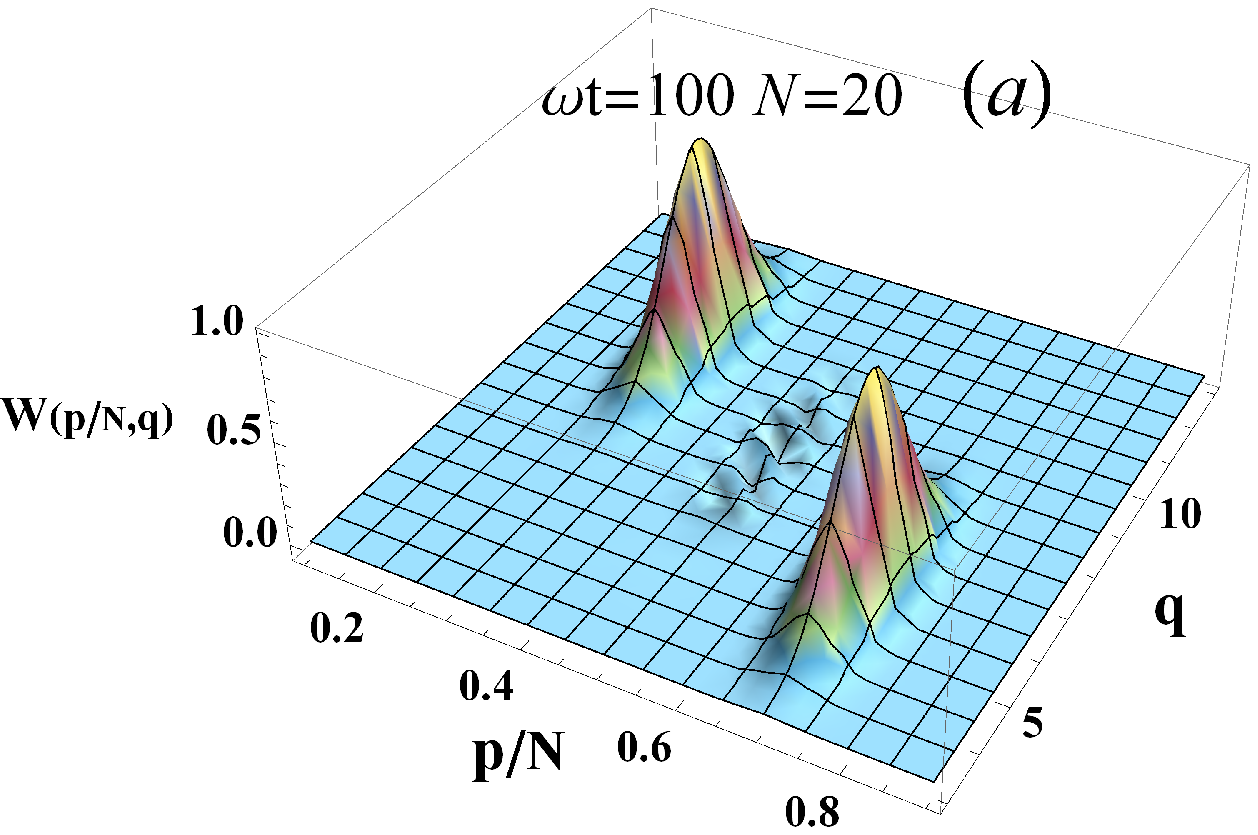}
\par\end{centering}
\begin{centering}
\includegraphics[scale=0.5]{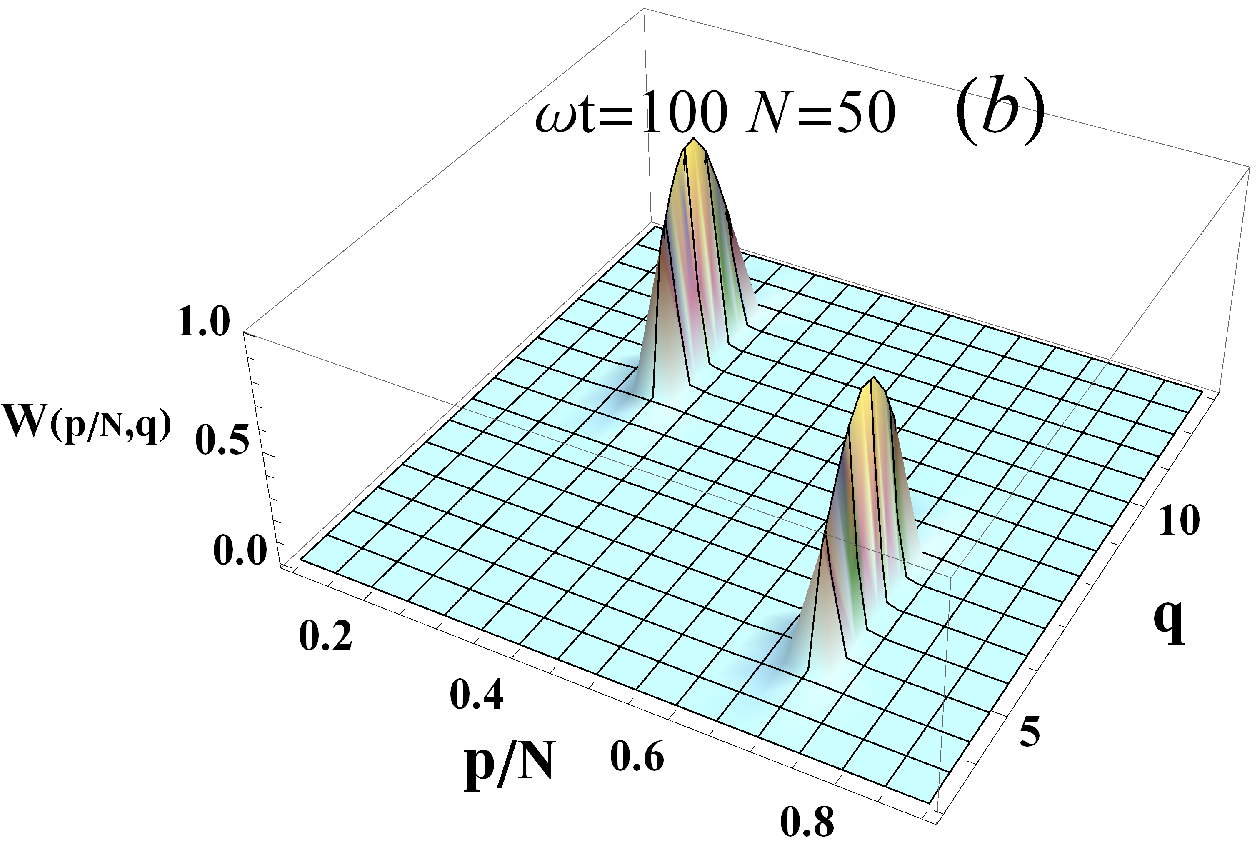}
\par\end{centering}
\caption{\label{fig:The-Wigner-function-N}The Wigner function in phase space
$(p/N,q)$ with different $N$.}
\end{figure}

When expanded at small time and large time limit, the above equation
becomes
\begin{align}
\lim_{\omega t\ll1}F(\alpha,\beta,t) & \propto\exp\left[-\frac{9N}{8}\omega^{2}t^{2}\left(\frac{\bigtriangleup E_{\alpha,\beta}}{Mc^{2}}\right)^{2}\right],\nonumber \\
\lim_{\omega t\gg1}F(\alpha,\beta,t) & \propto\exp\left[-\frac{9N}{32}\left(\frac{\bigtriangleup E_{\alpha,\beta}}{Mc^{2}}\right)^{2}\right],\label{eq:Fringe visibility function - two limits}
\end{align}
where $\bigtriangleup E_{\alpha,\beta}=\hbar^{2}\left(\Im(\alpha)^{2}-\Im(\beta)^{2}\right)/2M\sigma^{2}$
is the difference of mean kinetic energy of the two coherent states.
Fig. (\ref{fig:The-Wigner-function}) show the time evolution of the
Wigner function. When $t<1/\omega$ the interference term is damped
over time while when $\omega t=100\gg1$, a portion of coherence still
remains. Fig. (\ref{fig:The-Wigner-function-N}) show the Wigner function
at large time limit with different $N$ while the centers of the two
direct terms remain unchanged. And it is shown that the larger the
particle number $N$ is, the less coherence it keeps in large time
limit. All these features coincide with the analysis results given
in Eq. (\ref{eq:Fringe visibility function - two limits}). Now we
obtain the decoherence function of a macroscopic superposition quantum
state, $\sim\left|\alpha\right\rangle +\left|\beta\right\rangle $,
which depends on the mean momentum of the two superposed coherent
states $\hbar\Im(\alpha)/\sigma$ and $\hbar\Im(\beta)/\sigma$. What
is more, this function $F(\alpha,\beta,t)$ is highly similar as the
result we get in Sec. \ref{sec:Initial-condition,-time},
\begin{align}
\lim_{\omega t\ll1}\left|\rho_{c.m.}^{12}(t)\right| & \simeq\frac{1}{2}\exp\left[-\frac{9N}{8}\omega^{2}t^{2}\left(\frac{\bigtriangleup E_{1,2}}{Mc^{2}}\right)^{2}\right],\nonumber \\
\lim_{\omega t\gg1}\left|\rho_{c.m.}^{12}(t)\right| & \simeq\frac{1}{2}\exp\left[-\frac{9N}{32}\left(\frac{\bigtriangleup E_{1,2}}{Mc^{2}}\right)^{2}\right],
\end{align}
where $\bigtriangleup E_{1,2}$ is the difference of kinetic energy
of the two superposed states. Similar to the outcome in Sec. III,
in the superposition of coherent states case, the decoherence time
depends on the scale of the total system, the interaction strength
of real particles and the difference of the initial kinetic energy.

\section{Free-particles evolution}

In Sec. II and III, we studied decoherence of the CM in a ring with
N relativistic particles with the nearest-neighbouring interaction
and find that a restriction on particle number is necessary for the
decoherence of CM. Next, we will explore whether or not this particle
number restriction is induced by the nearest-neighbouring interaction.
To this end, in this section, we investigate the system of N free
relativistic particles
\[
H=\sum_{i}\sqrt{p_{i}^{2}c^{2}+m^{2}c^{4}}.
\]
We point out that the results about the cases without inter-coupling
could not be simply achieved from the above consequence by asumming
the couplings to vanish.

Making the Fourier and Bogoliubov transformation mentioned above in
Eqs. (\ref{eq:Fourier transformation}, \ref{eq:diagonalize}), one
find this Hamiltonian, to the second order approximation, becomes
\begin{equation}
H_{f}\simeq\frac{P^{2}}{2M}+\sum_{k=1}^{N-1}\frac{P_{k}^{2}}{2m}-\frac{3P^{2}}{2M^{2}c^{2}}\sum_{k=1}^{N-1}\frac{P_{k}^{2}}{2m},\label{eq:free particle 2-order}
\end{equation}
where $P$ describes the momentum of CM and $P_{k}$ the momentum
of $k$-th mode of relative motion.

As there are only momentum terms in Eq. (\ref{eq:free particle 2-order}),
the three terms commute pairwise. Therefore, for an initial state
such as the product state of the collective and relative momentum
operators, no decoherence will occur. What is more, in Sec. \ref{sec:Initial-condition,-time}
the initial state of the $k$-th mode relative motion is chosen to
be the ground state of the simple harmonic oscillator which in position
representation behaves as the Gaussian wave packet with width $\hbar/\left(4m\omega\sin\frac{\pi}{N}k\right)$.
To make sure that the two models start from the same condition, we
set the initial state of the free-particle model as
\[
\left|\psi(0)\right\rangle _{f}=\frac{\left|P_{1}\right\rangle +\left|P_{2}\right\rangle }{\sqrt{2}}\bigotimes\prod_{k}\left|\phi\right\rangle _{k},
\]
where $\left\langle x\mid\phi\right\rangle _{k}=\left(2\pi\sigma_{k}^{2}\right)^{-1/4}\exp\left[-x^{2}/4\sigma_{k}^{2}\right]$
is a Gaussian wave packet with packet width $\sigma_{k}^{2}=\hbar/\left(4m\omega\sin\frac{\pi}{N}k\right)$.
As the evolution of the state is governed by Eq. (\ref{eq:free particle 2-order}),
the state at time $t$ reads

\begin{widetext}

\begin{align}
\left|\psi(t)\right\rangle _{f} & =\frac{1}{\sqrt{2}}e^{-i\frac{P_{1}^{2}}{2\hbar M}t}\left|P_{1}\right\rangle \bigotimes\prod_{k}\int dpg_{k}(p)e^{-\frac{i}{\hbar}\left(1-\frac{3P_{1}^{2}}{2M^{2}c^{2}}\right)\frac{p^{2}}{2m}t}\left|p\right\rangle _{k}\nonumber \\
 & +\frac{1}{\sqrt{2}}e^{-i\frac{P_{2}^{2}}{2\hbar M}t}\left|P_{2}\right\rangle \bigotimes\prod_{k}\int dpg_{k}(p)e^{-\frac{i}{\hbar}\left(1-\frac{3P_{2}^{2}}{2M^{2}c^{2}}\right)\frac{p^{2}}{2m}t}\left|p\right\rangle _{k},
\end{align}

\end{widetext} where $g_{k}(p)=\left(2\sigma_{k}^{2}/\pi\hbar^{2}\right)^{1/4}e^{-p^{2}\sigma_{k}^{2}/\hbar^{2}}$.
Then the off-diagonal element of the reduced density matrix becomes
\begin{align*}
\left|\rho_{f}^{12}(t)\right| & =\frac{1}{2}\exp\left[-\frac{1}{4}\sum_{k=1}^{N-1}\ln\left(1+\frac{9\hbar^{2}}{16\sigma_{k}^{4}m^{2}}\left(\frac{\bigtriangleup E}{Mc^{2}}\right)^{2}t^{2}\right)\right].
\end{align*}

And for small $t$, we obtain
\begin{align}
\left|\rho_{f}^{12}(t)\right| & \simeq\frac{1}{2}\exp\left[-\frac{9}{8}N\omega^{2}t^{2}\left(\frac{\bigtriangleup E}{Mc^{2}}\right)^{2}\right],
\end{align}
with decoherence time
\begin{equation}
\tau_{f}=\frac{2\sqrt{2}Mc^{2}}{3\sqrt{N}\left|\triangle E_{1,2}\right|\omega}.
\end{equation}

Comparing these two model, we find that with the same initial state
the CM decoheres at the same rate as the outcome we obtained in Sec.
\ref{sec:Initial-condition,-time}, while there is no restriction
on the particle number ($N_{0}$) in the free-particle model. This
finding concludes that the restriction is introduced by the nearest-neighboring
interaction between relativistic particles in the ring. In fact, this
product state in the collective and relative movement reference frame
corresponds to an entanglement state in the real-particle movements
frame. Moreover, this entanglement state is difficult to prepare in
experiments since particles are all free. By the way, transformation
between the two frames considered in this paper is
\[
P_{k}=\begin{cases}
\sum_{j=1}^{N}\sqrt{\frac{2}{N}}p_{j}\cos\left(\frac{2\pi}{N}kj\right), & k\in[1,\frac{N-1}{2}]\\
\sum_{j=1}^{N}\sqrt{\frac{2}{N}}p_{j}\sin\left(\frac{2\pi}{N}kj\right), & k\in[\frac{N-1}{2},N-1]\\
\sum_{j=1}^{N}p_{j}, & k=N.
\end{cases}
\]

\section{Conclusions and discussions}

We have considered the relativistic modification in the Hamiltonian
of the N-particle-ring system with nearest-neighbouring interaction.
It is found that there exist interactions between the CM motion and
relative motion originated from the relativistic effect. As the part
of relative motions behaves as a harmonic oscillator bath environment,
this interaction causes the decoherence of the CM without dissipation.

Under the particle number condition, $N\gg N_{0}$, the decoherence
time of the system depends on the particle number of the system, the
coupling constant and the initial kinetic energy difference, i.e.,
$\tau\sim\left(3\sqrt{N}\left(\triangle E_{1,2}\right)\omega/2Mc^{2}\right)^{-1}$.
One can conclude that macroscopic objects decoherence faster than
microscopic ones. With a more classical state as the initial state,
the superposition of coherent states, the CM decoheres in a similar
way as the former case where the decoherence time depends on the two
expectation values of the CM momentum. Through a further study, we
find that the restriction of particle number $N_{0}$ is induced by
the nearest-neighbouring interaction of the ring. In the example we
take above, only in macroscopic systems with a particle number $N\sim N_{A}$,
the CM decoheres.

We finally remark that we only study the minimum decoherence mechanism
for the decoherence of MOs which is dipicted by its CM coupled with
relative movements due to relativistic effect. In real world, as we
said before, a physical system, especially a macroscopic system (with
lots of degrees of freedom) must interacts with its external environment
\cite{Zure-Rev-Mod-Phys,Joos-and-Zeh}. And the decoherence effect
caused by this external environment may dominate and the intrinsic
decoherence effect we considered here can be ignored. In other words,
in practice the MO will already be in a statistical mixture long before
reaching at the decoherence time we get in this paper.
\begin{acknowledgments}
One (G. H. Dong) of the authors thank Yao Yao for helpful discussions.
This work was supported by the National 973 program (Grant No. 2014CB921403),
the National Key Research and Development Program (Grant No. 2016YFA0301201),
the National Natural Science Foundation of China (Grant Nos.11421063
and 11534002) and NSAF (Grant No. U1530401).
\end{acknowledgments}

\appendix

\section{Squeeze operator in simple harmonic oscillator\label{sec:A}}

The Hamiltonian

\begin{align}
H & =\frac{p^{2}}{2m}+\frac{1}{2}m\omega^{2}x^{2}-\delta\frac{p^{2}}{2m}\nonumber \\
 & =\frac{p^{2}}{2m^{'}}+\frac{1}{2}m^{'}\omega^{'2}x^{2},
\end{align}
where $m^{'}=m/(1-\delta)$, $\omega^{'}=\omega\sqrt{1-\delta}$.
There are two kinds of definition of operators

\begin{align}
x & =\sqrt{\frac{\hbar}{2m\omega}}(a+a^{\dagger})\nonumber \\
p & =-i\sqrt{\frac{\hbar m\omega}{2}}(a-a^{\dagger})\nonumber \\
x & =\sqrt{\frac{\hbar}{2m^{'}\omega^{'}}}(b+b^{\dagger})\nonumber \\
p & =-i\sqrt{\frac{\hbar m^{'}\omega^{'}}{2}}(b-b^{\dagger}).
\end{align}

Then the Hamiltonian becomes
\begin{align}
H & =\hbar\omega\left(a^{\dagger}a+\frac{1}{2}\right)+\frac{\delta}{2}\frac{\hbar\omega}{2}(a-a^{\dagger})^{2}\nonumber \\
 & =\hbar\omega^{'}\left(b^{\dagger}b+\frac{1}{2}\right).
\end{align}

The eigenvalue of the system is $(n+1/2)\hbar\omega^{'}$. And

\begin{align}
b & =\sqrt{\frac{m^{'}\omega^{'}}{2\hbar}}\left(x+i\frac{p}{m^{'}\omega^{'}}\right)\nonumber \\
 & =\frac{1}{2}\left(\sqrt{\frac{m^{'}\omega^{'}}{m\omega}}(a+a^{\dagger})+\sqrt{\frac{m\omega}{m^{'}\omega^{'}}}(a-a^{\dagger})\right)\nonumber \\
 & =S^{\dagger}(r)aS(r),
\end{align}
where $r=\left|r\right|\exp(i\theta)$,$\left|r\right|=-\ln(1-\delta)/4$,
$\theta=\pi$ and $S(r)=\exp(r^{*}a^{2}/2-ra^{\dagger2}/2)$. Thus
\begin{equation}
H=\hbar\omega^{'}\left(S^{\dagger}(r)a^{\dagger}aS(r)+\frac{1}{2}\right).
\end{equation}

\section{Squeeze operator in simple harmonic oscillator\label{sec:B}}

\begin{widetext}
\begin{align*}
\left|\rho_{c.m.}^{12}(t)\right| & =\frac{1}{2}\prod_{k=1}^{N-1}\left|\left\langle 0\right|e^{i(H_{k}+H_{I,k}(P_{1}))t/\hbar}e^{-i(H_{k}+H_{I,k}(P_{2}))t/\hbar}\left|0\right\rangle _{k}\right|\\
 & =\frac{1}{2}\prod_{k=1}^{N-1}\left|f_{k}(P_{1},P_{2},t)\right|.
\end{align*}
where
\[
f_{k}(P_{1},P_{2},t)=\left\langle 0\right|S_{k}^{\dagger}(r_{1})e^{i\omega_{k}(P_{1})a^{\dagger}at}S_{k}(r_{1})S_{k}^{\dagger}(r_{2})e^{-i\omega_{k}(P_{2})a^{\dagger}at}S_{k}(r_{2})\left|0\right\rangle _{k},
\]
and
\begin{align*}
\omega_{k}(P)=\omega_{k}\sqrt{1-3P^{2}/2M^{2}c^{2}}, & S_{k}(r_{i})=\exp(-\left|r_{i}\right|a^{2}/2+\left|r_{i}\right|a^{\dagger2}/2),\left|r_{i}\right|=-\frac{1}{4}\ln(1-3P_{i}^{2}/2M^{2}c^{2}).
\end{align*}
Then we obtain

\begin{align*}
f_{k}(P_{1},P_{2},t) & =\left\langle 0\right|S_{k}(\xi_{1}e^{i\pi})S_{k}(r_{1})S_{k}(r_{2}e^{i\pi})S_{k}(\xi_{2})\left|0\right\rangle _{k}.
\end{align*}
where $\xi_{1}=\left|r_{1}\right|\exp\left[i(\pi-2\omega_{k}(P_{1})t)\right]$,
$\xi_{2}=\left|r_{2}\right|\exp\left[i(\pi-2\omega_{k}(P_{2})t)\right]$.
The squeeze operator can be transformed to \cite{Fisher squeeze operator 1984}
\begin{align*}
S(\left|z\right|e^{i\phi}) & =\exp(\left|z\right|e^{-i\phi}a^{2}/2-\left|z\right|e^{i\phi}a^{\dagger2}/2)\\
 & =\exp\left[-e^{i\phi}\tanh\left|z\right|L_{+}\right]\exp\left[-2\log\left(\cosh\left|z\right|\right)L_{3}\right]\exp\left[e^{-i\phi}\tanh\left|z\right|L_{-}\right],
\end{align*}
where $L_{+}=a^{\dagger2}/2$, $L_{-}=a^{2}/2$, and $L_{3}=\left(a^{\dagger}a+1/2\right)/2$
form a realization of the SU(l, l) Lie algebra.
\begin{align*}
f_{k}(P_{1},P_{2},t) & =\left\langle 0\right|S_{k}(\left|r_{1}\right|\exp\left[-2i\omega_{k}(P_{1})t)\right])S_{k}(\left|r_{1}\right|\exp\left[-i\pi)\right])S_{k}(\left|r_{2}\right|)S_{k}(\left|r_{2}\right|\exp\left[i(\pi-2\omega_{k}(P_{2})t)\right])\left|0\right\rangle _{k}\\
 & =\frac{1}{\sqrt{\cosh\left|r_{1}\right|\cosh\left|r_{2}\right|}}\left\langle 0\right|\exp\left[g_{1}^{k}L_{-}\right]\exp\left[\tanh\left|r_{1}\right|L_{+}\right]\exp\left[-2\log\left(\cosh\left|r_{1}\right|\right)L_{3}\right]\exp\left[-\tanh\left|r_{1}\right|L_{-}\right]\\
 & \cdot\exp\left[-\tanh\left|r_{2}\right|L_{+}\right]\exp\left[-2\log\left(\cosh\left|r_{2}\right|\right)L_{3}\right]\exp\left[\tanh\left|r_{2}\right|L_{-}\right]\exp\left[g_{2}^{k}L_{+}\right]\left|0\right\rangle _{k}.
\end{align*}
where $g_{1}^{k}=\exp\left(2i\omega_{k}(P_{1})t\right)\tanh\left|r_{1}\right|,$
$g_{2}^{k}=\exp\left(-2i\omega_{k}(P_{2})t\right)\tanh\left|r_{2}\right|$.
Following from inserting the identity operator $\int d^{2}\alpha/\pi\left|\alpha\right\rangle \left\langle \alpha\right|=\mathrm{I}$,
we obtain
\begin{align}
f_{k}(P_{1},P_{2},t) & =\frac{1}{\sqrt{\cosh\left|r_{1}\right|\cosh\left|r_{2}\right|}}\frac{1}{\pi^{3}}\int d^{2}\alpha_{1}...d^{2}\alpha_{3}\left\langle 0\right|\exp\left[g_{1}^{k}L_{-}\right]\left|\alpha_{1}\right\rangle \left\langle \alpha_{1}\right|\exp\left[\tanh\left|r_{1}\right|L_{+}\right]\nonumber \\
 & \exp\left[-2\log\left(\cosh\left|r_{1}\right|\right)L_{3}\right]\exp\left[-\tanh\left|r_{1}\right|L_{-}\right]\left|\alpha_{2}\right\rangle \left\langle \alpha_{2}\right|\exp\left[-\tanh\left|r_{2}\right|L_{+}\right]\nonumber \\
 & \exp\left[-2\log\left(\cosh\left|r_{2}\right|\right)L_{3}\right]\exp\left[\tanh\left|r_{2}\right|L_{-}\right]\left|\alpha_{3}\right\rangle \left\langle \alpha_{3}\right|\exp\left[g_{2}^{k}L_{+}\right]\left|0\right\rangle _{k}\nonumber \\
 & =\frac{1}{\cosh\left|r_{1}\right|\cosh\left|r_{2}\right|}\frac{1}{\pi^{3}}\int d^{2}\alpha_{1}...d^{2}\alpha_{3}\left\langle 0\right|\exp\left[\frac{g_{1}^{k}}{2}\alpha_{1}^{2}\right]\left|\alpha_{1}\right\rangle \left\langle \alpha_{1}\right|\exp\left[\frac{\tanh\left|r_{1}\right|}{2}\alpha_{1}^{*2}\right]\nonumber \\
 & \exp\left[-\log\left(\cosh\left|r_{1}\right|\right)a^{\dagger}a\right]\exp\left[-\frac{\tanh\left|r_{1}\right|}{2}\alpha_{2}^{2}\right]\left|\alpha_{2}\right\rangle \left\langle \alpha_{2}\right|\exp\left[-\frac{\tanh\left|r_{2}\right|}{2}\alpha_{2}^{2*}\right]\nonumber \\
 & \exp\left[-\log\left(\cosh\left|r_{2}\right|\right)a^{\dagger}a\right]\exp\left[\frac{\tanh\left|r_{2}\right|}{2}\alpha_{3}^{2}\right]\left|\alpha_{3}\right\rangle \left\langle \alpha_{3}\right|\exp\left[g_{2}^{k}L_{+}\right]\left|0\right\rangle _{k}.
\end{align}
As coherent state is over complete, the overlap of two different coherent
states is nonzero, i.e., $\left\langle \alpha\mid\beta\right\rangle =\exp\left[-\left(\left|\alpha\right|^{2}+\left|\beta\right|^{2}-2\alpha^{*}\beta\right)/2\right]$.
We also notice that coherent state is not the eigenstate of particle
number operator $a^{\dagger}a$, then
\begin{align*}
\exp\left[-\lambda a^{\dagger}a\right]\left|\alpha\right\rangle  & =e^{-\lambda a^{\dagger}a}e^{-\frac{\left|\alpha\right|^{2}}{2}}\sum_{n=0}^{\infty}\frac{\alpha^{n}}{\sqrt{n!}}\left|n\right\rangle =e^{-\frac{\left|\alpha\right|^{2}}{2}}e^{\left|\alpha e^{-\lambda}\right|^{2}/2}\left|\alpha e^{-\lambda}\right\rangle .
\end{align*}

In the following, we denote $\alpha_{i}=x_{2i-1}+ix_{2i}$, $\int d^{2}\alpha_{i}=\int_{-\infty}^{\infty}\int_{-\infty}^{\infty}dx_{2i-1}dx_{2i},(i=1,2,3).$
\begin{align}
f_{k}(P_{1},P_{2},t) & =\frac{1}{\cosh\left|r_{1}\right|\cosh\left|r_{2}\right|}\frac{1}{\pi^{3}}\int d^{2}\alpha_{1}...d^{2}\alpha_{3}\exp\left[\frac{g_{1}^{k}}{2}\alpha_{1}^{2}\right]\exp\left[-\frac{\left|\alpha_{1}\right|^{2}}{2}\right]\exp\left[\frac{\tanh\left|r_{1}\right|}{2}\alpha_{1}^{*2}\right]\nonumber \\
 & \exp\left[-\frac{\left|\alpha_{2}\right|^{2}}{2}-\frac{\left|\alpha_{1}\right|^{2}}{2}+\alpha_{1}^{*}\alpha_{2}e^{-\log\left(\cosh\left|r_{1}\right|\right)}\right]\exp\left[-\frac{\tanh\left|r_{1}\right|}{2}\alpha_{2}^{2}\right]\exp\left[-\frac{\tanh\left|r_{2}\right|}{2}\alpha_{2}^{2*}\right]\nonumber \\
 & \exp\left[-\frac{\left|\alpha_{2}\right|^{2}}{2}-\frac{\left|\alpha_{3}\right|^{2}}{2}+\alpha_{2}^{*}\alpha_{3}e^{-\log\left(\cosh\left|r_{2}\right|\right)}\right]\exp\left[\frac{\tanh\left|r_{2}\right|}{2}\alpha_{3}^{2}\right]\exp\left[-\frac{\left|\alpha_{3}\right|^{2}}{2}\right]\exp\left[\frac{g_{2}^{k}}{2}\alpha_{3}^{*2}\right]\nonumber \\
 & \exp\left[-x_{1}^{2}-x_{2}^{2}-x_{3}^{2}-x_{4}^{2}-x_{5}^{2}-x_{6}^{2}\right]\nonumber \\
 & =\frac{1}{\cosh\left|r_{1}\right|\cosh\left|r_{2}\right|}\frac{1}{\pi^{3}}\int dx_{1}...dx_{6}\exp\left[-\frac{1}{2}\sum_{i,j=1}^{6}A_{ij}^{k}x_{i}x_{j}\right],
\end{align}
where

\[
A^{k}(t)=\left(\begin{array}{ccc}
\Theta_{1} & \Omega_{1} & 0\\
\Omega_{1}^{T} & \Lambda_{1,2} & \Omega_{2}\\
0 & \Omega_{2}^{T} & \Theta_{2}
\end{array}\right),
\]
\begin{align*}
\Theta_{1} & =\left(\begin{array}{cc}
2-g_{1}^{k}-\tanh\left|r_{1}\right| & i\left(\tanh\left|r_{1}\right|-g_{1}^{k}\right)\\
i\left(\tanh\left|r_{1}\right|-g_{i}^{k}\right) & 2+g_{1}^{k}+\tanh\left|r_{1}\right|
\end{array}\right),\\
\Theta_{2} & =\left(\begin{array}{cc}
2-g_{2}^{k}-\tanh\left|r_{2}\right| & -i\left(\tanh\left|r_{2}\right|-g_{2}^{k}\right)\\
-i\left(\tanh\left|r_{2}\right|-g_{2}^{k}\right) & 2+g_{2}^{k}+\tanh\left|r_{2}\right|
\end{array}\right),\\
\Omega_{i} & =\left(\begin{array}{cc}
-\cosh^{-1}\left|r_{i}\right| & -i\cosh^{-1}\left|r_{i}\right|\\
i\cosh^{-1}\left|r_{i}\right| & -\cosh^{-1}\left|r_{i}\right|
\end{array}\right),\\
\Lambda_{1,2} & =\left(\begin{array}{cc}
2+\tanh\left|r_{1}\right|+\tanh\left|r_{2}\right| & i\left(\tanh\left|r_{1}\right|-\tanh\left|r_{2}\right|\right)\\
i\left(\tanh\left|r_{1}\right|-\tanh\left|r_{2}\right|\right) & 2-\tanh\left|r_{1}\right|-\tanh\left|r_{2}\right|
\end{array}\right).
\end{align*}
With the help of Gaussian integral \cite{path integral}, we obtain
\begin{align}
f_{k}(P_{1},P_{2},t) & =\frac{2^{3}}{\cosh\left|r_{1}\right|\cosh\left|r_{2}\right|}\frac{1}{\sqrt{\det\left[A^{k}(t)\right]}}=\frac{2^{3}}{\cosh\left|r_{1}\right|\cosh\left|r_{2}\right|}\frac{1}{\sqrt{\det\left[A^{k}(t)\right]}}.
\end{align}
In Eq. (\ref{eq:3-order couple neglected}), terms higher than $(p^{2})^{2}$
are neglected. Therefore the higher terms in $\left|\det\left[A^{k}(t)\right]\right|$
should also be ignored, i.e., $\left|r_{i}\right|\sim3P_{i}^{2}/8M^{2}c^{2}$,
$\cosh\left|r_{i}\right|\sim1+\left|r_{i}\right|^{2}/2$ and $\tanh\left|r_{1}\right|\sim\left|r_{i}\right|$.
After further calculation,
\begin{align}
 & \left|\cosh^{2}\left|r_{1}\right|\cosh^{2}\left|r_{2}\right|\det\left[A^{k}(t)\right]\right|\nonumber \\
 & \simeq\left|2^{6}\left(1+\left|r_{1}\right|^{2}+\left|r_{2}\right|^{2}\right)e^{-2i\omega_{k}(P_{2})t}\right|\nonumber \\
 & *\left|\left[\left|r_{1}\right|\left|r_{2}\right|\left(1-e^{2i\omega_{k}(P_{1})t}-e^{2i\omega_{k}(P_{2})t}+e^{2i\omega_{k}(P_{2})t}e^{2i\omega_{k}(P_{1})t}\right)-\left|r_{1}\right|^{2}e^{2i\omega_{k}(P_{1})t}e^{2i\omega_{k}(P_{2})t}-\left|r_{2}\right|^{2}+e^{2i\omega_{k}(P_{2})t}\right]\right|\nonumber \\
 & \simeq2^{6}\left|\left[1+\left|r_{1}\right|^{2}+\left|r_{2}\right|^{2}-\left|r_{1}\right|^{2}e^{2i\omega_{k}(P_{1})t}-\left|r_{2}\right|^{2}e^{-2i\omega_{k}(P_{2})t}+\left|r_{1}\right|\left|r_{2}\right|\left(e^{2i\omega_{k}(P_{1})t}+e^{-2i\omega_{k}(P_{2})t}-e^{-2i\omega_{k}(P_{2})t}e^{2i\omega_{k}(P_{1})t}-1\right)\right]\right|\nonumber \\
 & =2^{6}\left[1+\left(\left|r_{1}\right|-\left|r_{2}\right|\right)^{2}\left(1-\cos\left(2\omega_{k}t\right)\right)\right],
\end{align}
where we have assume $\omega_{k}(P_{i})\simeq\omega_{k}$.

\begin{align}
\left|\rho_{c.m.}^{12}(t)\right| & =\frac{1}{2}\prod_{k=1}^{N-1}\left|f_{k}(P_{1},P_{2},t)\right|\nonumber \\
 & =\frac{1}{2}\left(\prod_{k=1}^{N-1}2^{-6}\left|\cosh\left|r_{1}\right|\cosh\left|r_{2}\right|\det\left[A^{k}(t)\right]\right|\right)^{-1/2}\nonumber \\
 & \simeq\frac{1}{2}\prod_{k=1}^{N-1}\left[1-\frac{9}{32}\frac{\left(\triangle E_{1,2}\right)^{2}}{M^{2}c^{4}}\left(1-\cos\left(2\omega_{k}t\right)\right)\right].\label{eq:rho product}
\end{align}
There is a product of $(N-1)$ terms in Eq. (\ref{eq:rho product})
and we can take its logarithm,
\begin{align}
\ln2\left|\rho_{c.m.}^{12}(t)\right| & =\sum_{k=1}^{N-1}\ln\left[1-\frac{9}{32}\frac{\left(\triangle E_{1,2}\right)^{2}}{M^{2}c^{4}}\left(1-\cos\left(2\omega_{k}t\right)\right)\right]\nonumber \\
 & \simeq-\frac{9}{32}\frac{\left(\triangle E_{1,2}\right)^{2}}{M^{2}c^{4}}\sum_{k=1}^{N-1}\left(1-\cos\left(2\omega_{k}t\right)\right).
\end{align}

\end{widetext}

\end{document}